\documentclass[twocolumn,prb,citeautoscript,superscriptaddress,8pt]{revtex4-2}

\usepackage{graphicx}
\usepackage{gensymb}
\usepackage{color}
\usepackage[dvipsnames]{xcolor}
\usepackage{float}
\usepackage{upgreek}
\usepackage{bm}
\usepackage{amsmath,amssymb}
\usepackage{array}
\usepackage{hyperref}
\usepackage{booktabs}
\usepackage{multirow}
\newcolumntype{M}[1]{>{\centering\arraybackslash}m{#1}}

\hypersetup{urlcolor=teal, colorlinks=true, linkcolor=teal, citecolor=teal, linktocpage=true}

\begin{document}

\title{Optical spin defect pairs in cubic boron nitride}

\author{Josiah~E.~Hsi}
\affiliation{School of Science, RMIT University, Melbourne, VIC 3001, Australia}

\author{Islay~O.~Robertson}
\email{islay.robertson@rmit.edu.au}
\affiliation{School of Science, RMIT University, Melbourne, VIC 3001, Australia}

\author{Abhijit~Biswas}
\affiliation{Materials Science and NanoEngineering, Rice University, Houston, Texas 77005, United States}

\author{Jishnu~Murukeshan}
\affiliation{Materials Science and NanoEngineering, Rice University, Houston, Texas 77005, United States}

\author{Valery~Khabashesku}
\affiliation{Materials Science and NanoEngineering, Rice University, Houston, Texas 77005, United States}

\author{Alexander~J.~Healey}
\affiliation{School of Science, RMIT University, Melbourne, VIC 3001, Australia}

\author{Erin~S.~Grant}
\affiliation{School of Science, RMIT University, Melbourne, VIC 3001, Australia}

\author{David~A.~Broadway}
\affiliation{School of Science, RMIT University, Melbourne, VIC 3001, Australia}

\author{Mehran~Kianinia}
\affiliation{School of Mathematical and Physical Sciences, University of Technology Sydney, Ultimo, NSW 2007, Australia}
\affiliation{ARC Centre of Excellence for Transformative Meta-Optical Systems, Faculty of Science, University of Technology Sydney, Ultimo, NSW 2007, Australia}

\author{Igor~Aharonovich}
\affiliation{School of Mathematical and Physical Sciences, University of Technology Sydney, Ultimo, NSW 2007, Australia}
\affiliation{ARC Centre of Excellence for Transformative Meta-Optical Systems, Faculty of Science, University of Technology Sydney, Ultimo, NSW 2007, Australia}

\author{Pulickel~M.~Ajayan}
\affiliation{Materials Science and NanoEngineering, Rice University, Houston, Texas 77005, United States}

\author{Jean-Philippe~Tetienne}
\affiliation{School of Science, RMIT University, Melbourne, VIC 3001, Australia}

\begin{abstract}

Room-temperature optically active solid-state spin defects are widely known to be useful in quantum sensing applications, however, only a select range of materials have been found to host such systems. 
Recent measurements in the van der Waals material hexagonal boron nitride (hBN) have shown optically detected magnetic resonance (ODMR) with spin-1/2-like signatures can be explained by a charge transfer mechanism where charges move between adjacent defects forming weakly coupled spin pairs.
Interestingly, these ODMR signatures have been reported in a variety of materials aside from hBN, suggesting the spin pair model provides a potentially material agnostic approach for enabling ODMR. 
Here, we test whether the charge transfer mechanism is supported in a different crystal phase, and report on ODMR signatures in cubic boron nitride (cBN), showing all the characteristic properties identified in hBN are preserved.
We consider a selection of different cBN samples of varying size and observe ODMR from a single sub-micron cBN particle, paving the way towards sensing applications.
This work further expands understanding of the ubiquity of optical spin defect pairs, and establishes the potential for exploring quantum technologies with a wider range of materials.

\end{abstract}

\maketitle

Optically detected magnetic resonance (ODMR) of solid-state spin defects is the basis for a variety of experimental demonstrations for quantum applications including simulation~\cite{ejDavisNaturePhysics2023}, networking~\cite{slnHermansNature2022}, and sensing~\cite{ajHealeyNaturePhysics2022}.
The defining principles of ODMR are efficient optical spin initialisation and readout, enabled by an optical-spin interface, leading to simple and robust base protocols for spin control which in some cases are even operable at room-temperature~\cite{gWolfowiczNatureReviewsMaterials2021}.
A notable example is the negatively charged nitrogen-vacancy (NV) defect in diamond, which is one of the most studied room-temperature ODMR systems due to its large ODMR contrast, long spin coherence, and high optical brightness~\cite{mwDohertyPhysicsReports2013}.
While the diamond lattice is partly responsible for these properties, it remains technologically immature as a host material due to poor scalability and limited ability to integrate with other semiconductor materials. 
These shortcomings have led to an ongoing search for ODMR active solid-state spin defects in alternative host materials.

So far the most prominent examples of alternative defects and hosts have been silicon vacancy and divacancy defects in silicon carbide (SiC)~\cite{wfKoehlNature2011,mWidmannNatureMaterials2015}, and the negatively charged boron vacancy ($V_{\rm B}$) defect in hexagonal boron nitride (hBN)~\cite{aGottschollNatureMaterials2020,aGottschollScienceAdvances2021,aGottschollNatureCommunications2021}.
One of the reasons only a select few room-temperature defects have been identified is because specific conditions are required to create the necessary optical-spin interface. 
These systems, including the NV defect in diamond, rely on spin-dependent intersystem crossings between the ground and excited states~\cite{mAtatureNRM2018}.
More recently, a series of observations in hBN~\cite{nChejanovskyNatureMaterials2021,hlSternNatureCommunications2022,njGuoNatureCommunications2023,sScholtenNatureCommunications2023,hlSternNatureMaterials2024,xGaoNature2025,iRobertsonNaturePhysics2025,bWhitefieldNatureMaterials2026}, gallium nitride (GaN)~\cite{jLuoNatureMaterials2024}, and $\beta$-germanium disulphide ($\beta$-GeS$_2$)~\cite{wLiuNanoLetters2025,sVaidyaNanoLetters2025} have identified an ODMR signature akin to a $S=1/2$ resonance, which is generally considered to be inconsistent with the intersystem crossing optical-spin interface and may instead be better explained by spin dependent charge transfer between defect pairs~\cite{iRobertsonNaturePhysics2025}.
An interesting aspect of the charge transfer mechanism is the optical-spin interface is effectively separate from the specific material and defect properties, which significantly relaxes the conditions under which these systems may arise, \textit{i.e.}~the ODMR mechanism is agnostic to the host material, as evident from the variety of materials it has been observed in.
In this sense, the charge transfer mechanism has the potential to provide access to ODMR active defects in a wider array of materials than previously considered.
A simple way of exploring this possibility is to consider a different crystal phase of a material in which the charge transfer mechanism has already been observed.

Here we consider cubic boron nitride (cBN), which is a boron nitride polymorph with a diamond-like zincblende crystal structure~\cite{lVelMaterialsScienceandEngineeringB1991}.
While it is mostly employed as a high refractory material in mechanical applications due to its high hardness and thermal conductivity, much like hBN (which is well known to host a number of stable defects~\cite{lWestonPhysicalReviewB2018}), cBN is also a wide bandgap semiconductor and has been predicted to have similar levels of impurities~\cite{pPiquiniPhysicalReviewB1997,wOrellanaPhysicalReviewB2001,nlNguyen2024}.
In this regard, experimentally there have been reports of both paramagnetic defects~\cite{svNistorSolidStateCommunications2000,svNistorAppliedMagneticResonance2010}, and single emitters~\cite{rBuividasOpticsLetters2015,aTararanPhysicalReviewB2018,giLopezMoralesOpticalMaterialsExpress2020} in cBN, however, so far there has been no demonstration of ODMR.

In this work, we study ODMR signatures from defect ensembles in a series of cBN samples and find an overall consistency with previously reported measurements from defect ensembles in hBN.
Specifically, we investigate characteristic spin and optical properties which are indicators of the optical-spin defect pair model and the underlying charge transfer mechanism for ODMR, including: the magnetic field response of ODMR, the ubiquitous nature of the ODMR under different excitation wavelengths, and the spin and photodynamics~\cite{sScholtenNatureCommunications2023,pSinghAdvancedMaterials2025,iRobertsonNaturePhysics2025}.
In addition, we also explore the potential of future sensing applications using a combination of confocal and atomic force microscopy (AFM) techniques.
These results show the potential for the charge transfer mechanism to broaden the scope of host materials for ODMR active solid-state defects and establish cBN as a viable material for quantum sensing applications.

\begin{figure}[tb!]
    \centering
    \includegraphics{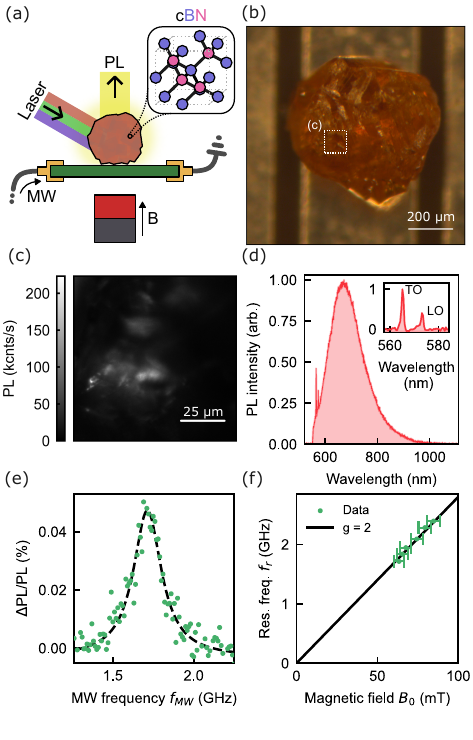}
    \caption{\textbf{Spin-active visible-band emitters in bulk cBN}. 
    (a)~Schematic of the experiment.  
    (b)~Microscope image of a sample of bulk cBN crystal. 
    (c)~Photoluminescence (PL) image of a surface of the cBN crystal, with bright and dark regions observed on the surface.
    (d)~PL spectrum of bright regions of cBN sample under 532 nm excitation. 
    Inset: A zoom in on the PL spectra showing Raman peaks at $565$ and $573$\,nm.
    (e)~ODMR spectrum fitted to a Lorentzian curve with peak contrast of $0.05$\% and FWHM of $0.21$\,GHz. 
    (f)~Resonant frequencies for a range of magnetic field strengths.
    Solid line indicates the free-electron resonance with $g = 2$. 
    Error bars correspond to a $5$\% uncertainty in the the magnetic field strength.
    }
    \label{fig:fig1}
\end{figure}

The experiment is depicted in Fig.~\ref{fig:fig1}(a).
Samples of cBN are placed on a printed circuit board (PCB) and illuminated with an excitation laser.
We first consider a large ($0.5$\,mm lateral size) crystal of cBN as shown in Fig.~\ref{fig:fig1}(b), the reddish colour suggesting a high impurity concentration in this sample.
Under $532$\,nm, excitation a weak photoluminescence (PL) signal is consistently observed across the sample, as are occasional locations of higher intensity~[Fig.~\ref{fig:fig1}(c)]. 
A PL spectrum taken from a region of the crystal including one of these bright spots yields a large peak centred at approximately $700$\,nm, with sharp, lower intensity features at $565$\,nm ($1103$\,cm$^{-1}$) and $573$\,nm ($1352$\,cm$^{-1}$)~[Fig.~\ref{fig:fig1}(d)].
The latter are associated with the cBN Raman lines (see SI) while we attribute the main broad peak to PL from an ensemble of optical emitters~\cite{tWerninghausAppliedPhysicsLetters1997}.
For a select bright region located in the cBN crystal, the ODMR spectrum reveals a single resonance peak with $0.05$\% contrast at $f_r\approx g\mu_{\rm B} B_0/h$ where $f_r$ is the resonance frequency, $B_0$ is the applied magnetic field strength, $g \approx 2$ is the Land{\'e} $g$-factor, $\mu_{\rm B}$ is the Bohr magneton, and $h$ is Planck's constant~[Fig.~\ref{fig:fig1}(e)].
The $g \approx 2$ scaling is confirmed by measuring the resonant frequency $f_r$ while varying the strength of the applied magnetic field, which was calibrated using a hBN sample as a reference~[Fig.~\ref{fig:fig1}(f)].
Both the PL and ODMR spectra bear resemblance to similar measurements of optical spin defect pairs in hBN samples.
Additionally, we note there is no consistent correlation between the optically bright regions and ODMR signal.
Instead ODMR is measured sporadically throughout the cBN sample in both bright and dim regions.

\begin{figure*}[tb!]
    \centering
    \includegraphics{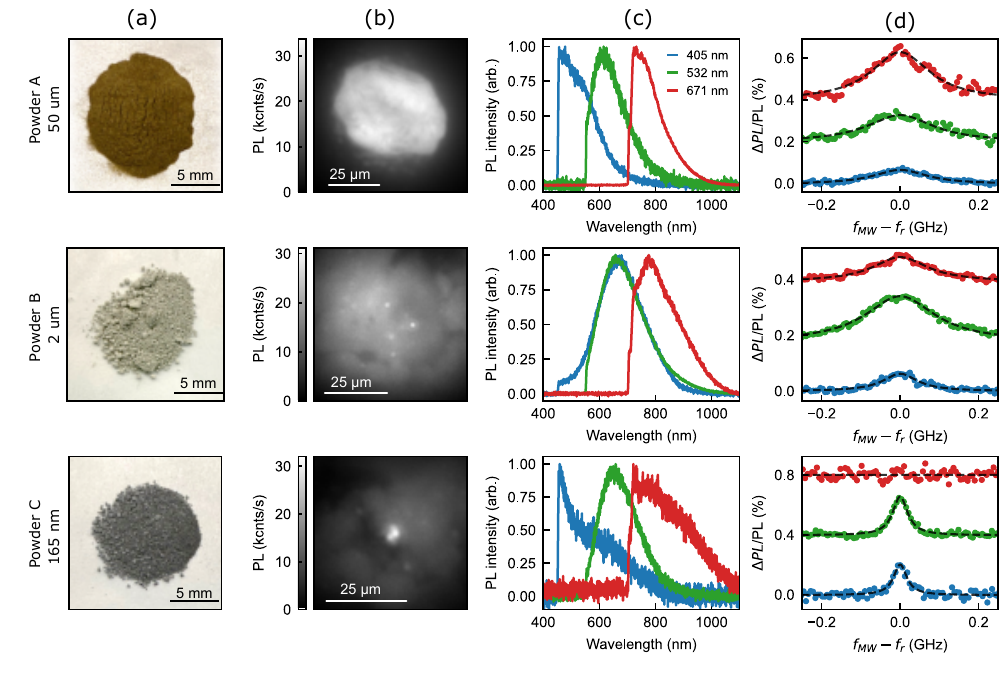}
    \caption{\textbf{Spin and optical measurements for cBN microparticles}. 
    (a)~Micrograph images of $50$\,$\mu$m, $2$\,$\mu$m, and $165$\,nm cBN grains. 
    (b)~Photoluminescence images from a single $50$\,$\mu$m grain, and from dense films of $2$\,$\mu$m and $165$\,nm particles, illuminated with a $532$\,nm laser.
    (c)~Photoluminescence spectra from the three cBN samples, illuminated with red ($671$\,nm), green ($532$\,nm), and blue ($405$\,nm) lasers. 
    (d)~ODMR spectra from cBN samples, illuminated with the red, green, and blue laser sources.
    Not all measurements were able to be taken under the same magnetic field conditions and so the spectra are plotted on a relative frequency axis $f_{\rm MW} - f_{\rm r}$\,GHz (See SI).
    }
    \label{fig:fig2}
\end{figure*}

To explore the effects of particle size we investigate three different commercially available samples of untreated (e.g. not subject to additional processing such as irradiation or annealing) micropowders with average grain sizes of $165$\,nm, $2$\,$\mu$m, and $50$\,$\mu$m~[Fig.~\ref{fig:fig2}(a)].
For measurement, we dilute the particles with isopropyl alcohol and then dropcast the resulting sediment directly on to a PCB.
Example images of the resultant PL under $532$\,nm excitation are shown in Fig.~\ref{fig:fig2}(b).
For the $50$\,$\mu$m particles, we find the PL intensity of individual grains varies considerably, including some that exhibit no PL response.
Unlike the large crystal shown in Fig.~\ref{fig:fig1}, the PL active particles show generally uniform PL throughout the entire grain of cBN.
The two smaller grain sizes show less intense, but more uniform PL across the aggregated particles with occasional bright spots distributed throughout.

A characteristic feature for the optical spin defect pairs in hBN is the ubiquity of the ODMR signal under a wide range of different wavelengths of laser excitation~\cite{pSinghAdvancedMaterials2025}.
This feature arises due to the interchangeability of the optical defect component in the optical spin defect pair model, allowing for multiple different optical emitters (\textit{e.g.}~with distinct absorption and emission spectra) to share similar ODMR signatures~\cite{iRobertsonNaturePhysics2025}.
To test whether cBN exhibits similar behaviour, PL spectra are measured for the three samples with $405$\,nm, $532$\,nm and $671$\,nm excitation and in all cases a single broad peak is observed ~[Fig.~\ref{fig:fig2}(b)].
In most cases, for the three samples, the peak intensity of the spectra falls within $\approx200$\,nm of the corresponding laser wavelengths with tails extending up to $\approx1000$\,nm.
This can be interpreted as the defect ensemble containing a distribution of zero-phonon lines which suggests the ensemble contains multiple distinct optical emitters~\cite{pSinghAdvancedMaterials2025}.
The only exception is for the $2$\,$\mu$m particles where the PL from both the $405$\,nm and $532$\,nm excitation lasers peak at $700$\,nm which instead suggests these lasers are exciting the same subset of emitters within the ensemble.

We next confirm the observed emitters contribute to the optical spin defect pair by performing ODMR measurements using the different excitation lasers~[Fig.~\ref{fig:fig2}(d)] (note, some measurements were taken using a pulsed sequence as this gave larger contrast).
Aside from the $165$\,nm grain sized powder under $671$\,nm illumination, a resonance was observed in all cases. 
Here, variations in the ODMR contrast cannot be attributed specifically to changes in excitation wavelength and may instead be the result of uncontrolled parameters, for example differences in the incident power density.

\begin{figure}[tb!]
    \centering
    \includegraphics{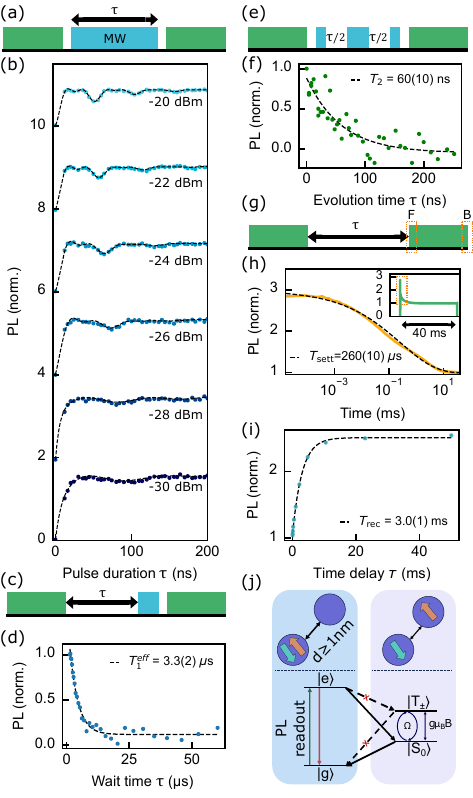}
    \caption{\textbf{Spin and photodynamic measurements}. 
    (a)~Rabi pulse sequence, with PL readout taken at the front of the $1$\,$\mu$s laser pulse after a MW pulse of duration $\tau$. 
    (b)~Rabi measurement of $165$\,nm powder sample at varying pre-amplified MW powers, fitted to a double exponential phenomenological model.
    (c)~Spin contrast decay measurement sequence.
    (d)~T$_1$ sequence measurement of $165$\,nm powder, fitted to a monoexponential.
    (e)~Hahn echo measurement pulse sequence
    (f)~T$_2$ Hahn echo measurement.
    (g)~Pulse sequence for PL settling-recovery measurements. Laser pulses are separated by a variable dark time $\tau$, with front (F) and back (B) of pulse PL read-out. 
    (h)~PL response at the front of the second laser pulse (F, inset) after wait time $\tau$. 
    The PL response is fitted with a stretched exponential to extract the settling time $T_{\rm sett}$.
    (i)~PL recovery rate as a function of dark time ($\tau$), calculated as F/B, along with a fitted stretched exponential.
    (j)~Spin-pair schematic in the single-particle picture and the associated many-body energy level structure. 
    }
    \label{fig:fig3}
\end{figure}

To verify the optical spin defect pair model we now probe the charge transfer mechanism by considering the spin and photodynamics of the cBN defect ensembles.
For these measurements, we focus on only the $165$\,nm cBN particles, as they are a more uniform source of PL and ODMR.
We first perform a Rabi experiment on the $165$\,nm particles~[Fig.~\ref{fig:fig3}(a)].
At lower MW powers, a low amplitude oscillation is observed which becomes more pronounced as the power is increased~[Fig.~\ref{fig:fig3}(b)].
Additionally, a secondary oscillation also emerges at higher powers.
We find the Rabi curves are well fit by a two frequency function with frequencies $\Omega_1$ and $\Omega_2$, where $\Omega_2$ is fixed by the relation $\Omega_2 = 2\Omega_1$, as expected from a weakly coupled spin pair system~\cite{dMcCameyPRL2010} (see SI).
Furthermore, at all driving powers the Rabi oscillations follow an asymmetric envelope, consistent with a metastable configuration of the spin pair which experiences spin dependent recombination back to its ground state~\cite{iRobertsonNaturePhysics2025}.

We further investigate the spin properties by probing the incoherent and coherent dynamics.
The sequence in Fig.~\ref{fig:fig3}(c) is a simple relaxation measurement to probe incoherent decay processes, where a $1$\,$\mu$s MW pulse is applied at the end of the dark time in a secondary sequence to mix the spin states for normalisation.
While this sequence is generally used to measure the longitudinal spin relaxation time $T_1$, the metastable configuration inferred from the Rabi measurements implies convolution with additional processes. 
As such, only an effective spin lifetime $T_1^{\rm eff}$ can be inferred, accounting for both spin-lattice and electronic relaxation.
By fitting the normalised relaxation curve with a monoexponential function we find $T_1^{\rm eff} \approx 3$\,$\mu$s, corresponding to the average spin polarisation lifetime of the defect ensemble contributing to the ODMR signal [Fig.~\ref{fig:fig3}(d)].
Subsequently, as coherent spin information is maintained for a few microseconds, we also measure the $T_2$ time using a normalised Hahn echo sequence shown in Fig.~\ref{fig:fig3}(e).
Fitting the resultant decay curve~[Fig.~\ref{fig:fig3}(f)] again with a monoexponential function we find $T_2 \approx 60$\,ns.

In addition to the spin dynamics, we also consider photodynamic measurements which grant further insight into the spin pair creation and recombination pathways.
These can be probed optically using the pulse sequence in Fig.~\ref{fig:fig3}(g), where we consider the PL response to long ($30$\,ms) laser pulses separated by a variable dark time $\tau$ which extends up to $50$\,ms.
We plot the timetrace of the entire PL response to the laser pulse after the maximum $\tau$ in Fig.~\ref{fig:fig3}(h).
The PL immediately increases to a maximum value before decaying to a steady state with a $1/e$ settling time $T_{\rm sett} \approx 260$\,$\mu$s which suggests optical pumping into a dark metastable state. 
Conversely, plotting the value of the PL overshoot (F/B, integrated over $5$\,$\mu$s of the pulse) as a function of $\tau$ shows a clear recovery of the PL, with a $1/e$ time $T_{\rm rec} \approx 3$\,ms as the system is left in the dark for longer periods indicating a recovery of the optical state as the metastable state decays [Fig.~\ref{fig:fig3}(i)].
Overall these values appear consistent with the broad range previously measured in hBN~\cite{iRobertsonNaturePhysics2025}.

In conjunction with the spin dynamics measurements, the photodynamic behaviour suggests the optically created metastable state is the weakly coupled spin pair which undergoes recombination in the dark.
This is consistent with the optical-spin interface based on charge transfer between optical spin defect pairs previously discovered in hBN [Fig.~\ref{fig:fig3}(j)].
The system manifests as two defects, one optically active defect, generally a spin singlet, and another remote defect.
When the system is in the ground state, two electrons are co-localised on the optical defect and upon optical excitation, a spin dependent charge transfer to the remote defect can occur which separates the two electrons forming the weakly coupled spin pair.
Quantum sensing is enabled by driving between the $T_{\pm}$ and $ST_0$ states of the weakly coupled spin pair which can then be optically read out after spin dependent recombination.

\begin{figure*}[tb!]
    \centering
    \includegraphics{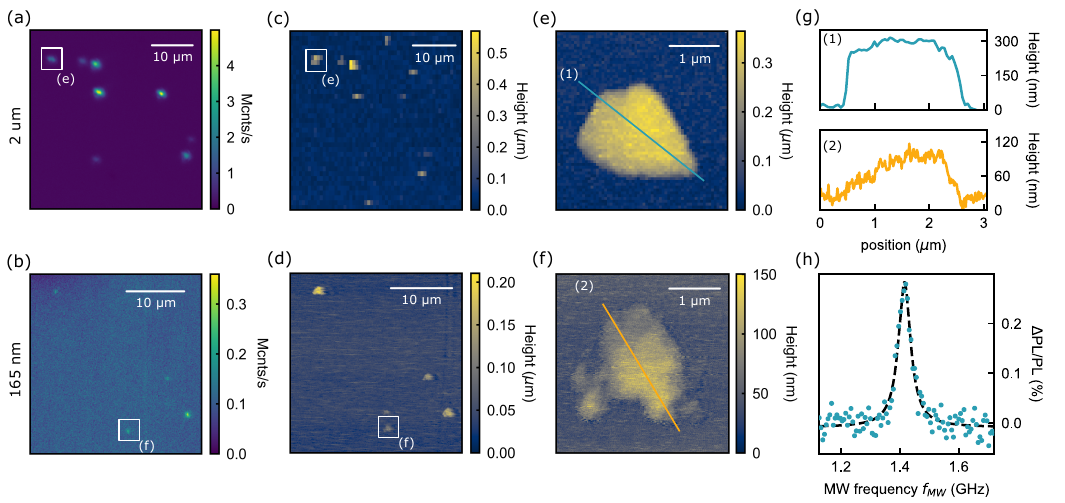}
    \caption{\textbf{Measurements of single cBN microparticles}.
    (a,b)~Confocal PL scan of a region of sparse 2 $\mu$m and 165 nm particles.
    (c)~Corresponding AFM scan of the same 2 $\mu$m particle region. Certain particles appear to be shifted relative to the PL scan.
    (d)~AFM scan of corresponding 165 nm particle region.
    (e)~AFM scan of a single PL-active 2 $\mu$m particle.
    (f)~AFM scan of a PL-active cluster of 165 nm particles
    (g)~Profile of 2 $\mu$m and 165 nm particles along the lines in (e) and (f). 
    (h)~ODMR from a single 2 $\mu$m particle on the same sample.
    }
    \label{fig:fig4}
\end{figure*}

Finally, we now explore the feasibility of micro and nanoscale sensing applications using cBN based optical spin defect pairs by isolating single particles from the $2$\,$\mu$m and $165$\,nm samples.
Particles of both sizes are dispersed by spin-coating dilute suspensions on to glass coverslips.
Using a confocal microscope, we map the position of fluorescent particles~[Fig.~\ref{fig:fig4}(a,b)], which we then correlate with topography images using an integrated AFM probe~[Fig.~\ref{fig:fig4}(c,d)].
On each sample, for a given field of view we observe a small number of particles.
From the PL maps, we are not able to distinguish single cBN particles from small aggregates, and so we perform additional AFM measurements on isolated regions to map the topography on the microscale~[Fig.~\ref{fig:fig4}(e,f)].
For the $2$\,$\mu$m sample we are able to measure a well-defined particle, while on the other hand for the $165$\,nm sample, the selected region appears to be more consistent with a particle aggregate.
Linecuts across the particle/aggregate, shown in Fig.~\ref{fig:fig4}(g), reveal the particle from the $2$\,$\mu$m sample has a plate-like shape, as exemplified in Fig.~\ref{fig:fig4}(e), where the particle height is $300$\,nm and the lateral size is $2$-$3$\,$\mu$m.
The maximum height of the aggregate is $\sim 100$\,nm which also suggests plate-like particles with a thickness less than the average size, similar to nanodiamonds~\cite{sEldemrdashCarbon2023}.
This picture is also consistent with SEM images (see SI).

One possible quantum sensing application of cBN particles is scanning magnetometry.
The simplest implementation of this technique involves attaching a particle embedded with quantum sensors on to an AFM tip.
While NV defects in diamond are commonly used for this purpose, this system requires precise alignment with an external bias field due to the inherent anisotropy of the $S=1$ ground state~\cite{jpTetienneNewJournalofPhysics2012}.
On the other hand, the spin-$1/2$-like nature of the weakly coupled spin pair system in cBN enables isotropic sensing and can thus simplify the experimental design of the scanning magnetometer~\cite{ioRobertsonACSPhotonics2025}, while also being more robust than hBN~\cite{xGaoNatureCommunications2024}.
In order to establish the viability of using cBN particles for this purpose, we measure ODMR from a defect ensemble in a single $2$\,$\mu$m particle~[Fig.~\ref{fig:fig4}(h)].
The contrast and linewidth from the single particle are consistent with the ODMR measurement from the powder films in Fig.~\ref{fig:fig3}, however, the size of the particle currently would inhibit the spatial resolution of a scanning magnetometry image. 
To realise the potential of cBN particles for scanning magnetometry, smaller particle sizes should be considered.
Taking an ODMR measurement from an isolated single $165$\,nm particle would require further understanding of the cBN surface chemistry in order to prevent aggregation.
Furthermore, the low ODMR contrast and short coherence time lead to an inferior sensitivity to the NV defect and thus further optimisation of the material and defect properties are required.

To summarise, we have observed ODMR from defect ensembles in a series of untreated commercial cBN samples. 
The ODMR signature exhibits the same features previously observed in hBN, notably a single ODMR resonance when under an applied magnetic field with spin-1/2-like scaling with magnetic field strength, which is observable under a wide range of excitation wavelengths.
Performing a series of spin and photodynamic measurements, we were able to associate these features with a defect ensemble containing multiple different optical emitters where ODMR arises from a charge transfer mechanism involving defect pairs.
Noting the structural differences between hBN and cBN, the consistency of the ODMR features implies the optical spin defect pair model based on the charge transfer mechanism is agnostic to crystal phase, and corroborates evidence from other works indicating such defect pair systems may be accessible in a diverse range of materials.
Further understanding of these defects would be enabled by investigating single emitters, which could lead to better insight into the defect structure through zero-phonon lines or hyperfine features in the ODMR to aid in \textit{ab initio} calculations.
Such studies may also reveal more exotic systems, such as recently identified spin complexes observed in hBN from the same class of defects~\cite{xGaoNature2025,bWhitefieldNatureMaterials2026}.
Finally, cBN based quantum sensors have the potential to enable sensing experiments in extreme conditions (e.g. $>800^\circ$C in air) currently out of reach of established material hosts for quantum sensors such as diamond, SiC, and hBN (which all burn or oxidise in air at elevated temperatures).

\section*{Acknowledgements}

This work was supported by the Australian Research Council (ARC) through grants
CE200100010, FT200100073, FT220100053, DE230100192, and DP250100973, and by the Office of Naval Research Global (N62909-22-1-2028). A portion of this work was supported by the U.S. Air Force Office of Scientific Research and Clarkson Aerospace Corp. under Award FA9550-24-1-0004.

\section*{Data availability}

The data supporting the findings of this study are available within the paper and its supplementary information files. 

\section*{Competing Interests Statement}

The authors declare no competing interests.

\appendix

\section{Experimental details} 

\subsection{Sample preparation} \label{sec:samplePrep}

We investigated four cBN samples in this work: large cBN crystals with $\approx 0.5$\,mm average diameter, and three cBN micro/nano particle samples of different sizes with $\approx 50$\,$\mu$m, $\approx 2$\,$\mu$m, and $\approx 165$\,nm average diameters.
The $165$\,nm powder was purchased from PlasmaChem GmbH, and the $2$\,$\mu$m and $65$\,$\mu$m were purchased from MSE Suppliers (Tucson, Arizona, USA).
All samples were studied as recieved without additional treatment (e.g. no irradiation, implantation, or thermal annealing).
For general optical and spin measurements, the cBN samples were either placed (large crystals) or drop cast (micro/nano particles) directly onto a printed circuit board (PCB) used for microwave delivery.
Samples were prepared for drop casting by mixing the particles in a larger volume of isopropyl alcohol (IPA) and then pipetted from the resulting sediment.
For the AFM measurements, the $65$\,$\mu$m and $2$\,$\mu$m partiles were diluted in water to $0.1$\,mg/mL and then spin coated at $3000$\,rpm onto a glass coverslip.

In addition to the cBN samples we also used a hBN nanopowder purchased in 2017 from Graphene Supermarket (BN Ultrafine Powder) as a reference for spin pair ODMR (see ref.~\cite{sScholtenNatureCommunications2023}). 
Similar to the cBN particle samples, the hBN nanopowder was drop cast directly onto a PCB used for microwave delivery.

\subsection{Experimental setup} \label{sec:expSetup}

The ensemble measurements were carried out on a custom-built wide-field fluorescence microscope. Optical excitation from a continuous-wave (CW) $\lambda = 532$\,nm laser (Laser Quantum Opus $2$\,W) was gated using an acousto-optic modulator (Gooch \& Housego R35085-5) and focused using a widefield lens to the back aperture of the objective lens (Nikon S Plan Fluor ELWD 20x, NA = 0.45). 
The photoluminescence (PL) from the sample was separated from the excitation light with a dichroic mirror and filtered using longpass and shortpass filters, before being sent to either (1) a scientific CMOS camera (Andor Zyla 5.5-W USB3) for imaging and ODMR measurements, or (2) an avalanche photodiode (Thorlabs APD410A) for time-resolved PL measurements, or (3) a spectrometer (Ocean Insight Maya2000-Pro) for PL spectroscopy.

Simultaneous confocal scanning and atomic force microscopy (AFM) was accomplished with a Mad City labs QS-PLL AFM integrated into a home-built confocal microscope. Akiyama probes in self-oscillation mode was used to acquire the AFM images. A 515 nm CW laser (Cobolt 06-MLD) was used for optical excitation, and focused onto the sample surface with an objective lens (Nikon S Plan Fluor ELWD 40x, NA = 0.6). Collected photoluminescence was filtered with a 515 nm longpass dichroic mirror and a separate 550 nm longpass filter, and focused onto a 62.5 $\mu$m core multimode optical fibre (ThorLabs GIF625) connected to an single photon counting module (Excelitas SPCM-AQRH-14-FC) or a spectrometer (Princeton Instruments Acton SpectraPro SP-2500).

\subsection{Material characterisation} \label{sec:materialCharac}

Without significant defining features in the spin and optical properties (e.g. hyperfine features in ODMR or zero-phonon lines) to distinguish the samples from hBN, we perform a series of material characterisation measurements to confirm the samples are cBN.

\begin{figure*}[tb!]
    \centering
    \includegraphics{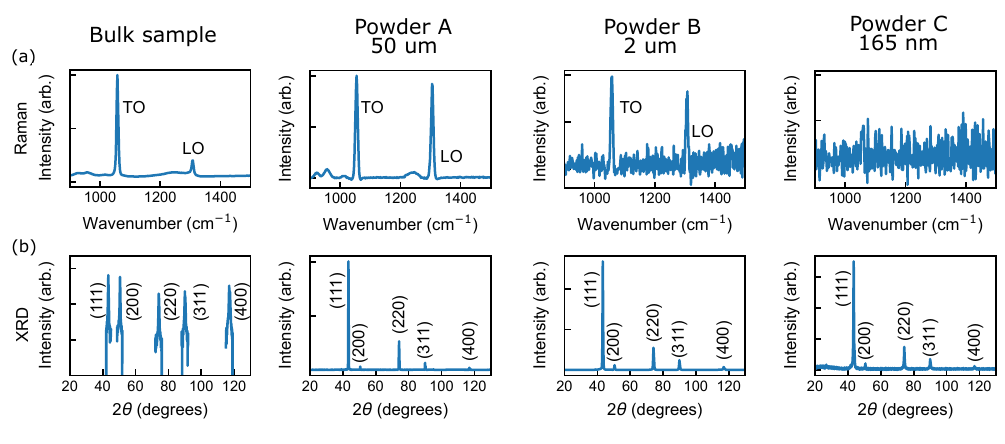}
    \caption{
    \textbf{Raman and XRD characterisation of cBN samples}.
    (a)~Raman spectra of all cBN samples used. TO and LO traces are visible in all but the smallest powder sample used.
    (b)~X-ray diffraction spectra of cBN samples. Individual (hkl) peak scans were performed for the bulk sample.
    }
    \label{fig:figA1}
\end{figure*}

Raman spectroscopy was conducted using a Renishaw inVia confocal microscope equipped with a $532$\,nm excitation laser [Fig.~\ref{fig:figA1}(a)]. 
Background subtraction was performed with the adaptive iteratively reweighted penalized least squares algorithm~\cite{zhangRSC2010}.
All samples except for the $165$\,nm show distinctive TO and LO peaks at $1054$\,cm$^{-1}$ and $1306$\,cm$^{-1}$ respectively. 
These values are commensurate with the separate measurements performed in main text Fig.~1(d) which were performed in a region of the bulk sample which also exhibits ODMR where we attribute the approximately $50$\,cm$^{-1}$ offset to a difference in spectrometer calibration.

X-ray diffraction (XRD) was performed using a Rigaku SmartLab thin-film diffractometer, with a monochromatic Cu K$\alpha$ radiation source ($\lambda = 1.5406$\,\AA), and a scan rate of $1^\circ$/min. 
For the bulk crystal, sample was oriented along the respective (hkl) facet and $2\theta$ scans were performed [Fig.~\ref{fig:figA1}(b)].
All samples exhibit identical spectra representative of the cBN lattice.

\begin{figure*}[tb!]
    \centering
    \includegraphics{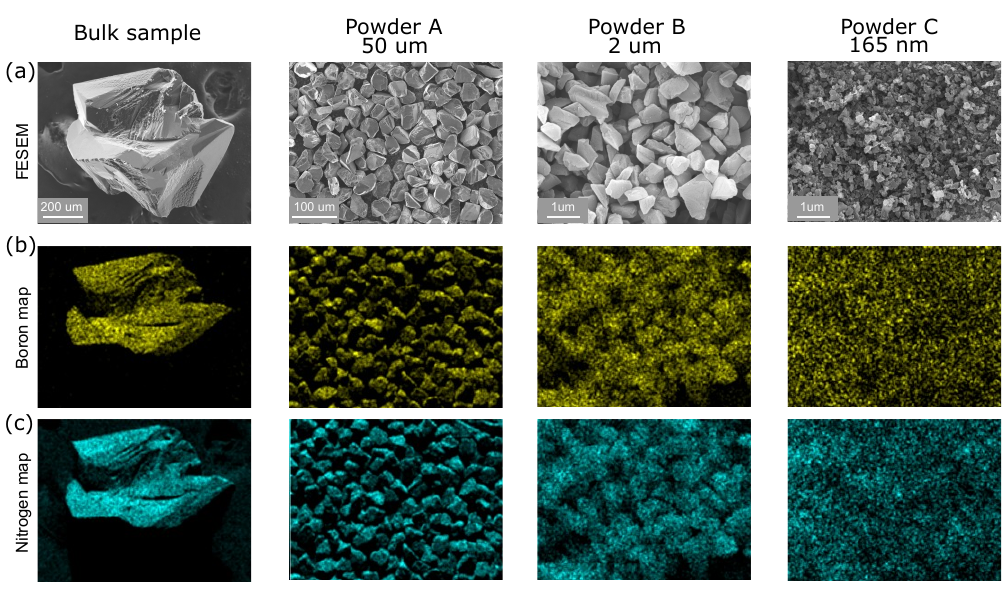}
    \caption{
    \textbf{FESEM and EDS scans of cBN samples}.
    (a)~Field emission scanning electron microscopy (FESEM) image of cBN samples.
    (b)~Boron image of the regions in (a) taken with energy-dispersive X-ray spectroscopy (EDS) mapping.
    (c)~Nitrogen image of the regions in (a) with EDS mapping.
    }
    \label{fig:figA2}
\end{figure*}

Surface morphology [Fig.~\ref{fig:figA2}(a)] and EDS mappings of the boron [Fig.~\ref{fig:figA2}(b)] and nitrogen [Fig.~\ref{fig:figA2}(c)] content were analyzed by field emission scanning electron microscopy (FESEM) using a JEOL JSM IT800 SHL FEG system using energy of $20$\,kV.
The FESEM images show  the powders consist of plate-like particles and confirm the average particle sizes.
The observed shapes and sizes are also in agreement with the AFM data in main text Fig.~4.
The EDS maps show high concentrations of boron and nitrogen in all samples.

\section{Rabi measurement analysis} 

\begin{figure}[tb!]
    \centering
    \includegraphics{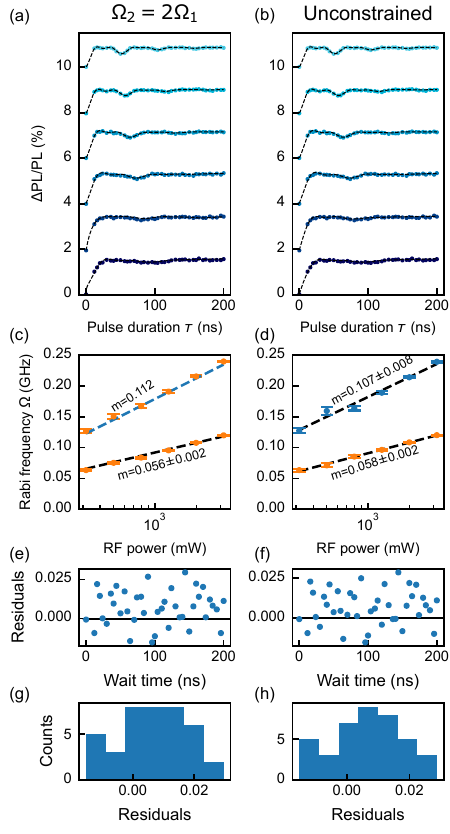}
    \caption{\textbf{Rabi frequency analysis}.
    (a)~Replicated from~Fig.~\ref{fig:fig3}b, Rabi measurements taken over a range of MW powers fit with a phenomenological model where the beat frequency is constrained to the Rabi frequency.
    (b)~Rabi measurement fit with unconstrained Rabi frequency components.
    (c)~Fitted Rabi frequency component $(\Omega_1)$, with linear fit to the components. 
    A second component equal to double the fitted linear slope, along with twice the fitted frequency components are also depicted for illustration.
    (d)~Unconstrained fitted Rabi frequency components along with linear fits.
    (e,f)~Residuals of fits to the Rabi oscillation model at -$20$\,dBm MW power for each Rabi frequency constraint.
    (g,h)~Histogram of residuals for each Rabi frequency constraint.
    }
    \label{fig:figA3}
\end{figure}

Here we rationalise our interpretation of the beat frequency in the Rabi oscillations as evidence for a weakly coupled spin pair system by showing explicitly the beat frequency scales at twice the Rabi frequency in both constrained and unconstrained fits of the data.
We fit the Rabi oscillations with the following phenomenological model\cite{sScholtenNatureCommunications2023},
\begin{equation*}
\begin{split}
    f(t)=&Ae^{-k_1t}\cos(\Omega_1 t)\\
    &+B(e^{-k_1t}+e^{k_2t})\cos(\Omega_2 t)
    +Ce^{-k_3t}+D
\end{split}
\label{eq:A1}
\end{equation*}
where the Rabi frequency $\Omega_1$ either constrains the beat frequency $\Omega_2$ with the relation $\Omega_1 = 2\Omega_2$ model~[Fig.~\ref{fig:figA3}(a)] or both are treated as independent variables~[Fig.~\ref{fig:figA3}(b)].
For both cases the data is well fit by the model and the Rabi frequency increases linearly when plotted against a logarithmically increasing MW power~[Fig.~\ref{fig:figA3}(c,d)]. 
Importantly, the extracted beat frequencies from the unconstrained fit match perfectly with the beat frequencies determined from the constrained fit. This is supported through residual analysis of both fits~[Fig.~\ref{fig:figA3}(e,f,g,h)] indicating no significant difference between constrained and unconstrained cases.
Confirming the $\Omega_1 = 2\Omega_2$ relationship strongly suggests the defect ensemble is composed of weakly coupled spin pairs.

\bibliography{references}

\end{document}